\documentclass[twocolumn,showpacs,preprintnumbers,amsmath,amssymb]{revtex4}

\usepackage{graphicx}
\usepackage{dcolumn}
\usepackage{bm}

\begin{document}

\preprint{}

\title{ High Magnetic Field Phase Diagram of PrOs$_4$Sb$_{12}$} 
\author{C.R. Rotundu, H. Tsujii, Y. Takano, and B. Andraka}%
\email{andraka@phys.ufl.edu}
\address{Department of Physics, University of Florida\\
P.O. Box 118440, Gainesville, Florida  32611-8440, USA}
\author{H. Sugawara, Y. Aoki, and H. Sato}
\address{Graduate School of Science, Tokyo Metropolitan University\\
Minami-Ohsawa 1-1, Hachioji, Tokyo 192-0397, Japan}

\date{\today}

\begin{abstract}

The magnetic phase diagram of PrOs$_4$Sb$_{12}$ has been investigated by specific heat measurements between 8 and 32 T. A new Schottky anomaly due to excitations between two lowest crystalline-electric-field (CEF) singlets, has been found for both $H \parallel (100)$ and $H \parallel (110)$ above the field where the field-induced ordered phase (FIOP) is suppressed. The constructed $H-T$ phase diagram shows weak magnetic anisotropy and implies a crossing of the two CEF levels at about 8 - 9 T for both field directions. These results provide an unambiguous evidence for the $\Gamma_1$ singlet being the CEF ground state and suggest the level crossing (involving lowest CEF levels) as the driving mechanism of FIOP.
\\
\end{abstract}

\bigskip
\maketitle

\narrowtext


A recent discovery of superconductivity\cite{Bauer} in PrOs$_4$Sb$_{12}$ is an important milestone in the field of strongly correlated electrons for a number reasons. Firstly, a large discontinuity in the specific heat at the superconducting transition temperature $T_c$ unambiguously establishes that Pr-based materials can support heavy fermion (hf) states. Secondly, proposed crystalline electric field (CEF) schemes suggest non-magnetic ground states and exclude a conventional Kondo effect, believed to be the source of hf behavior in Ce- and U-based metals. The only other serious microscopic mechanism considered as the origin of the hf behavior is a controversial quadrupolar Kondo effect. \cite{Cox}  Finally, two superconducting transitions observed\cite{Vollmer} in this compound imply an anisotropic order parameter.  Therefore, PrOs$_4$Sb$_{12}$ is a candidate for a novel mechanism of superconducting pairing that is due to neither phonons nor magnetic interactions.

Correct interpretations of these new exciting phenomena hinge crucially on the proper accounting for the CEF configuration of Pr. The consensus is that the non-magnetic (non-Kramers) $\Gamma_3$ doublet is the ground state, and  $\Gamma_5$ is the first excited triplet. However, available experimental data do not rule out the possibility of the $\Gamma_1$ singlet as the ground state.  Among the strongest arguments for both configurations were those provided by specific heat in zero and small magnetic fields. In particular, the zero field Schottky-like anomaly at 3.1 K can be related to the $\Gamma_3$ - $\Gamma_5$ model, assuming these two levels are split by 7.5 K, or $\Gamma_1$ - $\Gamma_5$ model with the splitting of 8.4 K. Inelastic neutron scattering measurements\cite{Maple} yield the lowest-energy peak at 8 K. Entropy changes in small fields and as a function of temperature were also used to argue for either of the models\cite{Bauer,Vollmer,Maple,Aoki}. The difficulty with interpreting these low temperature, low field results is related to strong hybridization of 4f and conduction electrons, as evident from the large electronic specific heat coefficient and the size of the discontinuity in the specific heat ($C$) at $T_c$. Therefore, we have performed specific heat measurements in high magnetic fields, in which a coupling between conduction and f electrons is expected to be suppressed, to get insight into the CEF configuration of Pr in PrOs$_4$Sb$_{12}$.  

The crystal used in our investigation was from the same batch of high quality samples for which a host of other measurements were reported.\cite{Aoki,dHvA} The specific heat study to 14 T was performed in a superconducting magnet. The measurements above 14 T were carried out in a resistive Bitter magnet. The calorimetry employed the relaxation method, in which a small sample of about 1.5 mg was used to optimize measuring conditions. Figs. 1 and 2 show specific heat in fields ranging from 10 to 32 T. The field was applied along the nominal (100) direction. Our estimate of the probable error in the sample alignment is about $5^o$. All the data presented are after subtracting the phonon background approximated by a Debye temperature ($\Theta_D$) of 165 K (after Vollmer et al.)\cite{Vollmer}). This value of $\Theta_D$ is quite controversial. In fact, Bauer et al.\cite{Bauer}, Aoki et al. \cite{Aoki}, and Maple et al.\cite{Maple} suggested $\Theta_D$ to be 304, 320, and 259 K, respectively. Changing $\Theta_D$ to 320 K does not alter the main conclusions of our study, but has some quantitative impact on the results, as discussed further. A relatively high low-temperature limit was chosen to avoid complications associated with a nuclear contribution of Pr. This nuclear contribution is strongly enhanced by coupling with orbital moments of f-electrons.\cite{nuclear} The specific heat, at temperatures where the nuclear degrees of freedom dominate, is difficult to measure by a conventional relaxation method because of an additional time scale(s) entering the experiment, the nuclear spin-lattice relaxation time $T_1$. \cite{AT} Strongly non-exponential temperature decays observed at the lowest temperatures (e.g., below 0.5 K in the field of 10 T and below 1.5 K in the field of 32 T) indicate the importance of nuclear degrees of freedom and cannot be analyzed using the so called $\tau_2$ correction. Therefore, these lowest temperature points carry large uncertainty and are not analyzed in detail.   
\begin{figure}[btp]
\begin{center}
\leavevmode 
\includegraphics[width=0.8\linewidth]{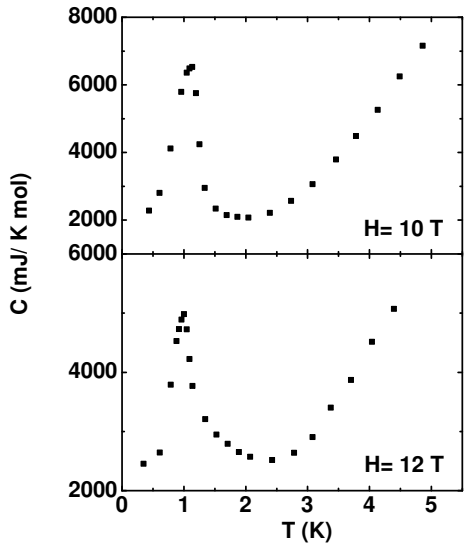} 
\caption{ Specific heat of PrOs$_4$Sb$_{12}$ in 10 and 12 T in the vicinity of FIOP transition. The magnetic field was applied along the (100) direction. Phonon contribution was subtracted assuming $\Theta_D$ =165 K.}
\label{fig1}
\end{center}
\end{figure}
\begin{figure}[btp]
\begin{center}
\leavevmode 
\includegraphics[width=0.8\linewidth]{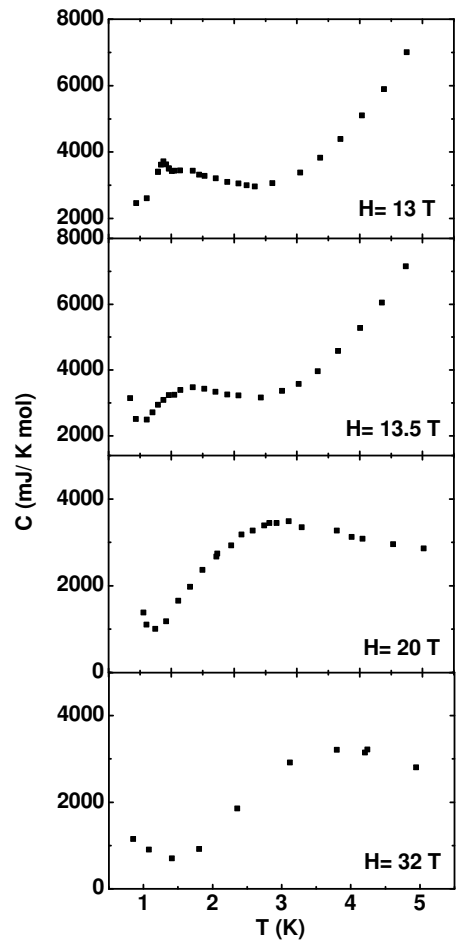} 
\caption{Specific heat of PrOs$_4$Sb$_{12}$ in magnetic fields 13, 13.5, 20, and 32 T applied parallel to (100), after phonon subtraction. Notice the appearance of a shoulder at about 1.2-1.3 K at 13 T and the disappearance of the FIOP transition at 13.5 T. The low temperature tails are due to nuclear contribution of Pr.}
\label{fig2}
\end{center}
\end{figure}

Figure 1 shows the low temperature specific heat for 10 and 12 T. Pronounced peaks in the vicinity of 1 K correspond to a transition to a field induced ordered phase (FIOP), which was first observed by Aoki et al.\cite{Aoki} for fields larger than 4 T also applied along the (100) direction. The highest field used in this latter study was 8 T. This long range order was confirmed by other investigations, including the specific heat of Vollmer et al.\cite{Vollmer} and magnetization study of Tayama et al.\cite {Tayama} It has been proposed that the order parameter corresponding to FIOP is antiferro-quadrupolar type. This type of order is consistent with large anomalies in the specific heat and a very small value of the ordered (antiferromagnetic) moment (about 0.025 $\mu$$_{B}$ at 0.25 K)\cite{Kohgi}. Combining our results with those of Aoki et al. \cite{Aoki} and Vollmer et al. \cite{Vollmer} shows that the transition temperature ($T_n$) identified as a position of the peak in the specific heat ($C$) reaches a maximum value somewhere near 9 T. Similarly, $C(T_n)$ has a non-monotonic field dependence,  attaining the highest value between 8 and 10 T. Notice a large reduction of $C(T_n)$ and small but clearly discernible decrease of $T_n$ between 10 and 12 T.  It is interesting to note a relatively large discrepancy for $C$ at $T_n$ between all three reports for overlapping fields. This discrepancy is probably due to the anisotropic response of the specific heat in magnetic fields near $T_n$, discussed below, and some misalignment of the sample with respect to the (100) direction. An increase of the field from 12 to 13 T (Fig. 2) results in a small reduction of $T_n$ and large suppression of $C(T_n)$. Moreover, a shoulder appears on the high temperature side of the FIOP anomaly. The specific heat value at this shoulder is about 3400 mJ/K mol. This shoulder evolves into a broad maximum for a slightly higher field of 13.5 T.  Again, the value of $C$ at the maximum is between 3400 and 3500 mJ/K mol.  In this field we no longer detect a signature corresponding to FIOP. Thus, our results strongly imply a disappearance of FIOP before $T_n$ reaches 0. A similarly broad peak exists at all fields studied up to at least 32 T. The temperature of this broad maximum increases with the field strength (Fig. 2).

The magnitude of the maximum, in fields of 13 T and larger, is essentially field independent and ranges from 3300 to 3500 mJ/K mol. These values are within about 10\% of the maximum value for a Schottky anomaly of a two level system with identical degeneracies. \cite{Gopal} The uncertainty of our specific heat measurements in these fields (and at temperatures where nuclear contribution is small) is about 10\%. In addition, there is the aforementioned uncertainty associated with a subtraction of the phonon term. Increasing the Debye temperature from 165 K, used in our subtraction, to the other extremal value proposed, 320 K, raises the estimate of the electronic part of $C$ by about 290 mJ/K mol at 3.5 K. Thus, the extracted values at the maximum are well within the realistic error bar of the theoretical 3650 mJ/K mol. Since the 32 T field is large enough to significantly split any degenerate levels, the observed Schottky anomaly has to represent excitations between two singlets. The temperature of the maximum, $T_m$, is related to the energy separation of the two levels $\delta$, $T_m=0.417\delta$. An extrapolation of $T_m$ to $T=0$ (Fig. 3) determines the field at which the two levels cross.  Our best estimate of this crossing field, using the lowest fields, is between 8 and 9 T.
\begin{figure}[btp]
\begin{center}
\leavevmode 
\includegraphics[width=0.8\linewidth]{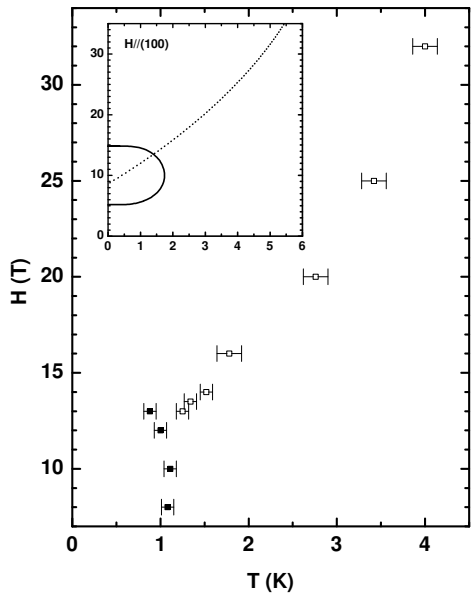}
\caption{ Magnetic field phase diagram of PrOs$_4$Sb$_{12}$ for the field parallel to (100). Filled squares represent the FIOP transition. Open squares correspond to the Schottky anomaly. The inset shows a model calculation of the H versus $T$ phase diagram assuming the singlet as the ground state. The solid line represents the FIOP boundary; the dashed line corresponds to a maximum in $C$ ($T_m=0.417\delta$; $\delta$ is the energy separation between the lowest CEF levels.) The model Hamiltonian and CEF parameters are identical to those of ref. 10.}
\label{fig3}
\end{center}
\end{figure}

These results can be used to infer new information pertaining to a plausible crystal field configuration of Pr. 
Pr can be modeled by the following single-site Hamiltonian\cite{Kohgi}. \begin{equation} 
\label{eq:1} 
   {\cal H} = {\cal H}_{\rm CEF} -g_J\mu_B{\bf J}{\bf H} -{\cal J} 
\langle{\bf J'}\rangle{\bf J} - \sum_{i}Q_i\langle O'_i\rangle O_i, 
\end{equation} 
where ${\cal H}_{\rm CEF}$, ${\bf J}$ and $O_i$ represent the CEF Hamiltonian for the cubic $T_h$ symmetry, the total angular momentum and the $i$-th quadrupole moment (see Ref.\cite{Kohgi} for detailed definition).
Note that the conventionally used ${\cal H}_{\rm CEF}$\cite{Bauer} with the $O_h$ symmetry\cite{Lea} is only approximate because it ignores the the tetrahedral symmetry of Sb ions.\cite{Takegahara}. This latter CEF Hamiltonian describes the Zeeman effect roughly but cannot account for the neutron diffraction results in the FIOP region\cite{Kohgi}. In particular, the lower $T_h$ symmetry is needed to explain the appearance of a small antiferromagnetic moment along the (010) direction, and not along (100), when the field is applied parallel to the (001) direction. (Directions (100) and (010) are not equivalent). Using the CEF parameters proposed by Kohgi\cite{Kohgi} for the $\Gamma_1$ - $\Gamma_5$ CEF configuration, $T_m$ (with $Q_{i}=0$) and the $O_{yz}$-type quadrupolar ordering temperature $T_n$ are calculated\cite{comment}. Note that this model consistently reproduces the inelastic neutron data\cite{Maple}, characteristic maxima appearing in the magnetic susceptibility\cite{Bauer,Aoki}, and the zero-field specific heat\cite{Bauer,Vollmer,Aoki}. The results of our calculations are presented in the inset to Fig. 3. 

The measured phase diagram (Fig. 3) and the theoretical one expected for the $\Gamma_1$ - $\Gamma_5$ model (inset to Fig. 3) are strikingly similar. In particular, in both diagrams, the crossing field is very close to the one at which the transition temperature of FIOP becomes maximum. Some overestimation of $T_n$ can be attributed to the mean-field nature of the calculations. On the other hand, there is less defined correlation between these two fields in the alternative $\Gamma_3$ - $\Gamma_5$ model. Instead, it has been argued for this latter model that FIOP is related to the crossing of the upper $\Gamma_3$ and lower $\Gamma_5$ levels.\cite{Vollmer} Thus, the observed correlation between the two characteristic fields constitutes our first argument for the $\Gamma_1$ singlet being the lowest CEF level.
\begin{figure}[btp]
\begin{center}
\leavevmode 
\includegraphics[width=0.8\linewidth]{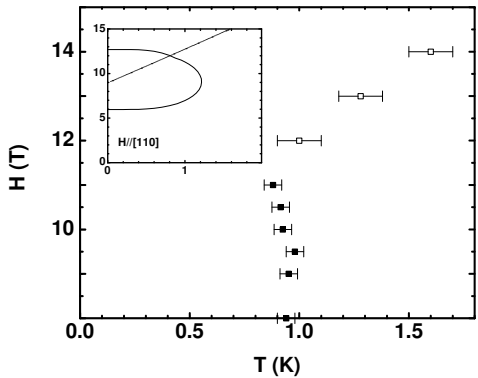}
\caption{ Magnetic field phase diagram of PrOs$_4$Sb$_{12}$ for $H$ parallel to (110). For the definition of symbols see Fig. 3.}
\label{fig4}
\end{center}
\end{figure}
 
More conclusive arguments regarding the CEF configuration can be obtained from the study of the anisotropy of the Zeeman effect. The Zeeman effect is expected to be strongly anisotropic for the $\Gamma_3$ doublet and almost isotropic for the $\Gamma_1$ singlet,\cite{Tayama} for the range of CEF parameters that are consistent with the susceptibility\cite{Bauer}. In particular, there is no crossing involving the lowest CEF levels when the field is applied along the (110) direction in the $\Gamma_3$ - $\Gamma_5$ model. We emphasize that this prediction is quite general; it does not depend on whether one uses the the CEF Hamiltonians and parametrs proposed by Bauer et al.\cite{Bauer} or Kohgi et al.\cite{Kohgi}. Figure 4 shows the measured high field phase diagram for this field direction. For this direction we observe a decrease of $T_n$ values with respect to the (100) direction for the corresponding fields, consistent with the previous magnetization measurements\cite{Tayama}. On the other hand, within the uncertainty of the measurement, there is no change in the position of the Schottky anomaly at 13 and 14 T, as expected for the $\Gamma_1$ CEF ground state and inconsistent with the $\Gamma_3$ scenario. Moreover, for the (110) orientation we clearly observe the Schottky anomaly already at 12 T. This lower field limit for the Schottky maximum is probably due to competition between the two types of anomalies, discussed below, and lower values of $T_n$ for the (110) direction. A straight line fit for the three $T_m$ points results in the crossing field value of 8.5 $\pm$ 0.5 T. This value agrees with our estimate for the (100) direction.

Existence of the crossing field for the (110) direction provides an unambiguous evidence for the $\Gamma_1$ - $\Gamma_5$ model, in fact the strongest evidence to date. We stress that a small misalignment of the sample with respect to the field in either of the measurements cannot explain essentially identical crossing fields for both directions. In fact, the measured difference in $T_n$ values for (100) and (110) directions provides an additional check of the alignment. Similarly to the (100) direction, there seems to be a close correlation between the crossing field and the field corresponding to $T_n$ maximum. 
 
Figures 3 and 4 imply a strong competition between the field-induced order and Schottky peak. The FIOP transition in the specific heat abruptly disappears before $T_n$ reaches zero. Precise magnetization\cite{Tayama} and de Haas van Alphen measurements\cite{dHvA}, on the other hand, were able to map $T_n$ as a function of the magnetic field all the way to $T_n \approx 0$. This apparent contradiction can be explained by a very small entropy available for the FIOP transition above 13 and 12 T for fields parallel to the (100) and (110) directions, respectively. Specific heat, being a bulk measurement, can be less sensitive than magnetization techniques in this situation. A strong competition is to be expected in the $\Gamma_1$ - $\Gamma_5$ scenario. The ground state pseudo-doublet formed at the level crossing carries both magnetic and quadrupolar moments. Since a quadrupolar moment operator does not commute with a dipolar one, the quadrupolar interactions leading to FIOP compete with the magnetic Zeeman effect. 

Summarizing, we find an evidence for the level crossing occurring around 8-9 T. This field value is very close to the field at which both $T_n$ and $C(T_n)$ become maximum, suggesting that the crossing of the lowest CEF levels is the driving mechanism of FIOP. In fields above 10 T there is a strong competition between FIOP and the Schottky peak. Our high field specific heat results provide compelling arguments for the $\Gamma_1$ singlet being the CEF ground state of Pr in this system. This results call for a new model of electronic mass enhancement, instead of the ordinary quadrupolar Kondo effect, which has been discussed mainly in the context of the $\Gamma_3$ non-Kramers doublet ground state.    
 
This work has been supported by the U.S. Department of Energy, Grant No.
DE-FG02-99ER45748, National Science Foundation, DMR-0104240, the National High Magnetic Field Laboratory, and the Grant-in-Aid for Scientific Research from MEXT of Japan. We thank G.R. Stewart, P. Kumar, K. Ingersent, and R. Shiina for stimulating discussions.



\begin{thebibliography}{14}
\bibitem{comment}
For simplicity, we label CEF states by symbols used in the $O_h$ symmetry. Refer to Ref. 13, for the corresponding labels in the $T_h$ symmetry (e.g. $\Gamma_3 \to \Gamma_{23}$).
\end{thebibliography}
\end{document}